\documentclass{article}
\begin{document}
\title{Exterior Differential System for Cosmological $G_2$ Perfect Fluids 
and Geodesic Completeness }
\author{L. Fern\'andez-Jambrina\\
Departamento de Ense\~nanzas B\'asicas de la Ingenier\'{\i}a Naval\\
E.T.S.I. Navales\\ 
Arco de la Victoria s/n\\
E-28040-Madrid \\
and \\
L. M. Gonz\'alez-Romero\\
Departamento de F\'{\i}sica Te\'orica II\\
Facultad de Ciencias F\'{\i}sicas\\ 
Universidad Complutense\\
E-28040-Madrid}
\date{}

\maketitle
\begin{abstract}
In this paper a new formalism based on exterior differential systems is derived 
 for perfect-fluid spacetimes endowed with an abelian orthogonally transitive
$G_2$ group of motions acting on spacelike surfaces. This formulation 
allows simplifications of Einstein equations and it can be applied for
different purposes. As an example a singularity-free metric is rederived in
this framework. A sufficient condition for a diagonal metric to be geodesically
complete is also provided.
\end{abstract}

\section{Introduction}

Perfect-fluid spacetimes endowed with an abelian $G_2$ group of isometries have
been used for describing many different physical situations. When the group is
acting on timelike surfaces, they have been extensively used for describing
axisymmetric compact objects in stationary rotation (cfr. \cite{Esc} for a
review). 

On the other hand, if the group acts on spacelike surfaces, 
the applications are different \cite{Kramer}.  
A classification of these spacetimes is given in
\cite{class}.  They can
model spacetimes where two plane gravitational waves are colliding (cfr.
\cite{wave} for a review), but they are also useful for describing inhomogeneous cosmologies (cfr.
\cite{Corn},\cite{kra} for a review) in an attempt of coping with the
inhomogeneity present  in our Universe. An interesting feature of
$G_2$ cosmologies is the possibility of avoiding initial and final singularities (cfr. \cite{Esc},
\cite{seno}, \cite{grg} for a review) and therefore physics will be valid in the
whole spacetime. These models satisfy the causality and energy conditions and
just fail to contain trapped sets, according to the well-known singularity
theorems
\cite{HE}, \cite{Beem}.

In this paper we shall try to cope with two features concerning nonsingular
perfect fluid orthogonally transitive $G_2$ cosmological models: First of all
one has to devise a method for obtanining exact models and then one should
check whether the solution is singular or not. Both of these will be the aim
of this paper.

A new tetrad formalism based on differential forms will be introduced for
deriving results about spatially inhomogeneous spacetimes. It will be shown that
the methods initially devised for stationary axially symmetric spacetimes
\cite{ch} are also useful when the group of motions acts on spacelike surfaces.
The $1$-forms that are used in this formalism will be shown to have kinematical
meaning when considering spatial congruences. If the congruence corresponds
to an invariantly defined quantity (for instance, in the cylindrical case, the
axial Killing), the kinematical properties can be used  to classify the
solutions. Also, using the remaining gauge freedom, the tetrad can be adapted to
the congruence so that the exterior system can be written in terms of
$1$-forms with an invariant and physical interpretation. Moreover, when one
tries to obtain an exact model, it is generally useful to impose certain
assumptions on these kinematical quantities (now with an invariant
interpretation). As the approach is grounded on an exterior
differential system, the coordinates can be chosen according to the Ansatz that
is performed, instead of fixing them from the beginning.

Assuming one has obtained a new cosmological model, it usually takes lengthy
calculations to determine if there are singularitities in it. The question is
settled if the curvature invariants are already singular, but if they are
regular in the whole spacetime, there is a priori no reason to assume that
every geodesic is complete. In fact, there are cosmologies with regular
curvature invariants that are incomplete and therefore singular, despite no
quantity becomes unboundedly large \cite{HE}. Therefore we deem convenient to
have a general result that may simplify the task of analyzing the issue of
geodesic completeness. This matter will be addressed for the diagonal case in
this paper.

Let us describe in more detail the contents of the paper:

 In the second section spacelike congruences in a general spacetime will be
studied in order to achieve an interpretation for the mathematical quantities
that will appear in the formalism. In the stationary axisymmetric case timelike
congruences were considered and their tangent field could be considered as a
velocity and therefore the interpretation was straightforward. Another way of
interpreting spacelike congruences is due to Greenberg \cite{Greenberg}, but in
this paper a different approach will be followed.

In the third section the formalism is written in terms of an exterior system of
equations that include Cartan and Einstein field equations as well as their
integrability conditions. The set of equations will be simplified taking
advantage of the remaining gauge freedom.

As an example of how the exterior system can be used for obtaining exact solutions the
singularity-free model in \cite{leo} will be obtained within the formalism in
 the fourth section.

The question of geodesic completeness of diagonal inhomogeneous cosmological models will be
addressed in the fifth section and a theorem will be derived as a sufficient
condition for a model to be nonsingular. 

This condition will be shown to be weak enough to comprise all known diagonal singularity-free
inhomogeneous cosmological models in the sixth section.

\section{Spacelike congruences}

The kinematical properties of a timelike congruence can be defined by 
decomposing the covariant derivative of the timelike vector field 
defined by the congruence \cite{Ehlers}, \cite{Ellis}. An analysis of the 
spacelike congruences has been done by Greenberg \cite{Greenberg}. In 
his analysis Greenberg introduces an observer moving with a four 
velocity $w^a$. Projecting the covariant derivative of the 
vector field defined by the congruence orthogonal to $w^a$
 he obtains ``spacelike'' quantities characterizing the congruence. 
In this section we will follow a different approach to study the 
kinematical properties of a spacelike congruence. We will use a
straightforward translation of the timelike congruence analysis. In 
this way we obtain kinematical  properties not necessarily with a 
spacelike character, but we will find that these kinematical 
quantities can be used in a natural way to formulate a simplified 
differential form approach   for spacetimes with 
two commuting spacelike Killing vectors.

Let us consider a congruence of spacelike curves,
$$x^{\alpha} = x^{\alpha} (y^{a},\tau) \;\; , \;\; a = 1,2,3,$$
where the three constants $y^{a}= c^{a}$ specify a particular 
curve and $\tau$ is the arc-length. We can define a unit tangent 
vector,
$$n^{\beta} = \frac{d x^{\beta}}{d \tau} \;\; , \; \; 
n_{\alpha}n^{\alpha} = 1 ,$$

We will assume in the following that the congruence defines, at 
least locally, a vector field $n^{\beta}(x^{\alpha}$).

Let us take a particular curve $(C)$ in the congruence, specified by three 
constants 
($y^{a}$), and a point $p$ in the curve $(C)$ characterized by an 
arc-length $\tau$. Now consider another curve in the 
congruence $(C^{*})$ near $(C)$ and 
specified by constants $y^{a *} = y^a + \delta y^a$ and a point 
$p^{*}$ in $(C^{*})$ with the same arc-length ($\tau$) as $p$. Then, 
up to first order, we have,
$$\delta x^{\alpha} = \frac{\partial x^{\alpha}}{\partial y^a} \delta 
y^a,$$
which, in general, is not orthogonal to $n^{\alpha}$. In order  
to get a vector orthogonal to
$n^{\alpha}$ we introduce the projector  tensor, 
$$P^{\alpha}_{\beta} = g^{\alpha}_{\beta} - n^{\alpha} n_{\beta},$$
such that $P^{\alpha}_{\beta} n^{\beta} = 0$. Then we can define,
$$\delta_{\perp} x^{\alpha} = P^{\alpha}_{\beta} \delta x^{\beta} \;\; \;\; 
( \delta_{\perp} x^{\alpha} n_{\alpha} = 0 ) .$$ 
Note that $\delta_{\perp} x^{\alpha}$ can be spacelike, timelike, or 
null. 

The rate of change of the connecting vector of two spacelike curves of 
the congruences allows us to characterize locally the congruence. 
First, let 
us consider  the change in the `modulus'. Assume 
$\delta_{\perp} x^{\alpha}$ is timelike or spacelike, then,
$$ \epsilon (\delta l)^2 = g_{\alpha \beta} \delta_{\perp} x^{\alpha} 
\delta_{\perp} x^{\beta} \;\; , \;\; \epsilon = \pm 1 .$$

The rate of change of $\delta l$ along the congruence is,
\begin{equation}
\frac{(\delta l)\dot{\ }}{\delta l} = \epsilon S_{\alpha \beta} 
\frac{\delta_{\perp} x^{\alpha}}{\delta l} 
\frac{\delta_{\perp} x^{\beta}}{\delta l} + \frac{1}{3} \Phi,
\label{rateexp}
\end{equation}
where we define,
\begin{eqnarray}
     \Phi & \equiv & n^{\alpha}_{\ ;\alpha}
	\label{flux}  \\
	\dot{n}_{\alpha} & \equiv & n_{\alpha ; \beta} n^{\beta}
	\label{curv} \\
	S_{\alpha \beta} & \equiv & n_{(\alpha ; \beta)}-\dot{n}_{(\alpha} n_{\beta)} - 
	\frac{1}{3} \Phi P_{\alpha \beta},
	\label{def} 
\end{eqnarray}
where $S$ is a trace-free symmetric tensor.

If we define the unitary connecting vector,
$$e^{\alpha} \equiv \frac{\delta_{\perp} x^{\alpha}}{\delta l}$$
the equation (\ref{rateexp}) can be written as follows,
\begin{equation}
\frac{(\delta l)\dot{\ }}{\delta l} = \epsilon S_{\alpha \beta} 
e^{\alpha} e^{\beta}
 + \frac{1}{3} \Phi .
\label{rateexp2}
\end{equation}

On the other hand for the rate of change of $e^{\alpha}$ we get,
\begin{equation}
	P^{\beta}_{\ \alpha} (e^{\alpha})\dot{\ } = (W^{\beta}_{\ \alpha} + 
	S^{\beta}_{\ \alpha} - \epsilon S_{\mu \nu} e^{\mu} e^{\nu} 
	\delta^{\beta}_{\ \alpha}) e^{\alpha}
	\label{ratevect}
\end{equation}
where,
$$
W_{\alpha \beta} \equiv n_{[\alpha ; \beta]} - 
\dot{n}_{[\alpha} n_{\beta]}. 
$$ 

If $e^{\alpha}$ is an eigenvector of $S_{\mu \nu}$ ($S^{\mu}_{\ \nu} 
e^{\nu} =  \lambda e^{\mu}$) then the previous equations reduce to,
 \begin{equation}
 P^{\beta}_{\ \alpha} (e^{\alpha})\dot{\ } = W^{\beta}_{\ \alpha} e^{\alpha}.	
 	\label{rvectred}
 \end{equation}
 
 It is interesting to note that if we know $\dot{n}_{\alpha}$, 
 $S_{\alpha \beta}$, $W_{\alpha \beta}$, and $\Phi$ in a given point 
 we can reconstruct the congruence locally using that,
 $$n_{\alpha ; \beta} = \dot{n}_{\alpha} n_{\beta} + W_{\alpha \beta} +
 S_{\alpha \beta} + \frac{1}{3} \Phi P_{\alpha \beta}.$$
 
 Some important properties are,
 $$P_{\alpha \beta} n^{\beta} = 0 \; , \;
\dot{n}_{\alpha} n^{\alpha} = 0 \; , \; 
W_{\alpha \beta} n^{\beta} = 0 = S_{\alpha \beta} n^{\beta}. $$

In order to have a better characterization of the spacelike 
congruence we can study the eigenvalue problem for the trace-free symmetric 
three-dimensional tensor $S_{\alpha \beta}$ that can be formulated as
follows,
$$(S_{\alpha \beta} - \lambda P_{\alpha \beta}) 
v^{\beta} = 0.$$

As the quadratic forms defined by $S_{\alpha \beta}$ and $P_{\alpha \beta}$ are
not 
 definite forms, a standard analysis gives us the 
following different situations:
\begin{itemize}
	\item  $\lambda_{1}$, $\lambda_{2}$, and
$\lambda_{3}=-\lambda_{1}-\lambda_{2}$ 
	are three different real numbers. In this case we have one timelike 
	eigenvector and two spacelike eigenvectors. They are mutually 
	orthogonal.

	\item  $\lambda_{1}$, $\bar{\lambda_{1}}$ are complex conjugate and 
	$\lambda_{2}=-2\,\textrm{Re}(\lambda_{1})$. In this case we have two
complex conjugate 
	eigenvectors, 
	$m$ and $\bar{m}$ corresponding to the complex eigenvalues and a 
	real spacelike eigenvector orthogonal to $m$ and $\bar{m}$. 
	The complex eigenvector can be written 
	as $m = a + i b$, where $a$ is timelike and $b$ spacelike 
	and they are mutually orthogonal and  normalized to $-1/2$ and $1/2$ 
	respectively.

	\item  $\lambda_{1} = \lambda_{2}\neq 0$,  $\lambda_{3}=-2\,\lambda_{1}$ are
real numbers.
	In this case there is a null eigenvector corresponding to the double 
	eigenvalue and one spacelike eigenvector orthogonal to the plane 
	containing the null eigenvector. 

	\item  $\lambda_{1} = \lambda_{2} = \lambda_{3}= 0$. In this case there is a
null eigenvector corresponding to the triple eigenvalue.
\end{itemize}

The interpretation of the kinematical properties of a spacelike 
congruence depends on the physical interpretation of the vector field 
$n$
and the character of the  eigenvalues an 
eigenvectors of 
$S_{\alpha \beta}$. Assume for instance that the three eingenvalues are 
real and different. As a consequence the three eigenvectors are 
mutually orthogonal and there is one timelike and two spacelike 
eigenvectors. 
In order to give an interpretation in this case,
let us consider a curve (C) in the congruence and a point (P) in it 
corresponding to $\tau = \tau_{0}$. Assume that in the spacelike 
subspace in the
 orthogonal hyperspace to (C) in (P) we have a disk. Orthogonal to 
 it there is a unit timelike vector. If we imagine   different 
 curves of the congruences crossing the 
 border of the disk and the final point of the timelike vector, 
going from $\tau_{0}$ to 
 $\tau_{0} + \delta \tau $, the disk changes to an ellipse and the 
 direction and modulus of the timelike vector also changes. These changes 
 are determined by the kinematical variables of the congruences $\Phi$, 
 $S_{\alpha \beta}$, and $W_{\alpha \beta} $ and 
 the final effect can be decomposed in the following two steps: 
 
 \noindent {\bf Step 1: ($\Phi$ and $S_{\alpha \beta}$ effects)}
 
 The influence of $\Phi$ and $S_{\alpha \beta}$ on the disk and 
 timelike vector can be understood using  equation 
 (\ref{rateexp2}). If $e^{\alpha}$ is an eigenvector of $S_{\alpha 
 \beta}$ and verifies  $S_{\alpha \beta} e^{\beta} = \lambda e^{\alpha}$
 the equation reduces to,
 $$ \frac{(\delta l)\dot{\ }}{\delta l} = \lambda + \frac{1}{3} \Phi .$$
 Hence,  the  two directions indicated by the two spacelike 
 eigenvectors will transform  to the principal directions of the ellipse and 
 the length of the principal axis will be obtained from the previous 
 equation. The new direction of the timelike vector will be given by the 
 direction of the timelike eigenvector and the modulus will be 
 obtained  from the previous equation. In this way we can think of the 
 spacelike congruence as generating a ``virtual tube flux'' where the 
 transverse section is given by the ellipse, the velocity of the fluid 
 by the timelike unit eigenvector of $S_{\alpha \beta}$ and 
 its modulus can be interpreted as the density, whose changes are 
 given by the previous equation (note 
 that, in principle, the virtual fluid has nothing to do with any real  perfect 
 fluid. The physical interpretation of such virtual fluid depends of the 
 physical interpretation of the congruence).   
   
 \noindent {\bf Step 2: ($W_{\alpha \beta}$ effects) }
 
Finally the effect of $W_{\alpha \beta}$  on the ``tube flux'' is to 
``bend 
and twist the tube''. This result follows from the equation (\ref{rvectred}).

The only remaining kinematical variable  is $\dot{n}_{\alpha}$ but 
this represents the curvature  of the  selected curve in the 
congruence.

In the case when $S_{\alpha \beta}$ has degenerate or complex 
eigenvalues the interpretation is not so straightforward. We will 
study those cases in the particular situation when the metric admits 
two Killing vectors. 

\subsection{Kinematical Variables in a tetrad formalism}
Using a tetrad adapted to $n$ such that $\theta^2 \equiv n$,
$$ds^2 = -\theta^0 \otimes \theta^0 + \theta^1 \otimes \theta^1 +
\theta^2 \otimes \theta^2 + \theta^3 \otimes \theta^3, $$
the covariant derivative of $n$ takes the following form,
$$n_{a;b} = - \gamma_{2ab},$$
where $\gamma^a\, _{bc}=-\theta^a_{i;j}e^i_be^j_c$ (where $\{e_a\}$ is the
orthonormal frame dual to $\{\theta^a\}$) are the Ricci rotation coefficients.
The  kinematical properties of the congruence of
$n$ read as follows,
 \begin{eqnarray}
 	\dot{n}  & = & \gamma_{022} \theta^0 + \gamma_{122} \theta^1 + 
 	\gamma_{322} \theta^3
 	\label{acel}  \\
 	\Phi & = & -\gamma_{020} + \gamma_{121} - \gamma_{233}
 	\label{fluxi}  \\
 	W_{T} & =  & (\gamma_{021}-\gamma_{120}) \theta^0 \wedge 
 	\theta^1 + (\gamma_{023} + \gamma_{230}) \theta^0 \wedge \theta^3 +
 	(\gamma_{123} + \gamma_{231}) \theta^1 \wedge \theta^3
 	\label{rotdob}  \\
	S_{T} & = & (\frac{2}{3} \gamma_{020} + \frac{1}{3} \gamma_{121} - 
	\frac{1}{3} \gamma_{233}) \theta^0 \otimes \theta^0 + 
	 (\frac{1}{3} \gamma_{020} + \frac{2}{3} \gamma_{121} + 
	\frac{1}{3} \gamma_{233}) \theta^1 \otimes \theta^1 + 
	\nonumber\\
	& & (\frac{1}{3} \gamma_{020} - \frac{1}{3} \gamma_{121} - 
	\frac{2}{3} \gamma_{233}) \theta^3 \otimes \theta^3 + 
	\label{defor} \\
	& & (\gamma_{021} + \gamma_{120}) \theta^0 \otimes_s \theta^1 +
	(\gamma_{023} - \gamma_{230}) \theta^0 \otimes_s \theta^3 + 
	(\gamma_{123} - \gamma_{231)}) \theta^1 \otimes_s \theta^3,
	\nonumber
\end{eqnarray}
where $\otimes_s$ is the symmetric part of the tensor product, that is,
$v\otimes_s w=(v\otimes w+w\otimes v)/2$.

\section{Tetrad formalism for a spacetime endowed with two spacelike 
 orthogonally transitive commuting Killing vectors}
If a spacetime has two spacelike Killing vectors $\{\xi , \eta\}$ 
defining an abelian orthogonally transitive $G_{2}$ group of motions 
  we can choose the tetrad for the spacetime 
$\{\theta^{0}, \theta^{1}, \theta^{2} ,\theta^{3} \}$,
$$ds^2 = -\theta^0 \otimes \theta^0 + \theta^1 \otimes \theta^1 +
\theta^2 \otimes \theta^2 + \theta^3 \otimes \theta^3, $$
such that $\theta^{2}$ and $\theta^{3}$ are in ${\rm lin} \{\xi, \eta\}$ and
 \cite{Javier}, \cite{ch}. 
$$L_{\xi} \theta^{0} = 0= L_{\eta} \theta^{0} \; ; \; 
L_{\xi} \theta^{1} = 0= L_{\eta} \theta^{1},$$ 
$$L_{\xi} \theta^{2} = 0= L_{\eta} \theta^{2} \; ; \; 
L_{\xi} \theta^{3} = 0= L_{\eta} \theta^{3}.$$
It is important to notice that there is a residual $SO(1,1)$ gauge in 
the $\{\theta^{0}-\theta^{1} \}$ subspace and also a $SO(2)$ gauge in 
the $\{\theta^{2}-\theta^{3} \}$ subspace.

For the tetrad presented above  a family of independent 
non-vanishing Ricci rotation coefficients are the following,
$$\gamma_{010}, \gamma_{011}, \gamma_{022}, \gamma_{023}= \gamma_{032}, 
\gamma_{033}, \gamma_{122}, \gamma_{123}= \gamma_{132},
 \gamma_{133}, \gamma_{230}, \gamma_{231}.$$

Then, the kinematical variables reduce to,
  \begin{eqnarray}
 	\dot{n}  & = & \gamma_{022} \theta^0 + \gamma_{122} \theta^1 
 	\label{acels}  \\
 	\Phi & = & 0
 	\label{fluxis}  \\
    W_{T} & =  &  (\gamma_{023} + \gamma_{230}) \theta^0 \wedge \theta^3 +
 	(\gamma_{123} + \gamma_{231}) \theta^1 \wedge \theta^3
 	\label{rotdobs}  \\
	S_{T} & = & (\gamma_{023} - \gamma_{230}) \theta^0 \otimes_s \theta^3 + 
	(\gamma_{123} - \gamma_{231)}) \theta^1 \otimes_s \theta^3
	\label{defors}
\end{eqnarray}

It is interesting to define 1-forms $\alpha$, $\omega$, and $\sigma$,
\begin{eqnarray}
 \alpha & = & - \gamma_{022} \theta^0 - \gamma_{122} \theta^1
 	\label{K1}  \\
 \omega	 & = & - (\gamma_{023} + \gamma_{230}) \theta^0 - 
 (\gamma_{123} + \gamma_{231}) \theta^1
 	\label{K2}  \\
 \sigma	 & = & (\gamma_{230} - \gamma_{023}) \theta^0 + 
 (\gamma_{231} - \gamma_{123}) \theta^1
 	\label{K3}  
 \end{eqnarray}
 such that the kinematical variables can be written as,
 \begin{eqnarray*}
  	\dot{n} & = & - \alpha \\
  	W_{T} & = & - \omega \wedge \theta^3  \\
  	S_{T} & = & - \sigma \otimes_s \theta^3
  \end{eqnarray*} 
so that $\alpha$, $\omega$, and $\sigma$ completely parametrize the 
kinematical variables. 

The eigenvalue problem for $S_{T}$ simplifies and there are 
  three different cases. Defining $A \equiv 
\gamma_{023}-\gamma_{230}$ and $B \equiv \gamma_{123}-\gamma_{231}$ 
we get:
\begin{enumerate}
	\item  $B = \epsilon A $ where $\epsilon^2 = 1$ ($\sigma$ null)
	
	In this case $\sigma$ is a null form 
	$\sigma	= A (\theta^0 + \epsilon \theta^1).$ 
		We have $\lambda_{1} = \lambda_{2} = \lambda_{3} = 0$ whose 
		eigenvector is proportional to $\sigma$.

	\item  $B > A$ ($\sigma$ spacelike)
	
	In this case we have  
	$\lambda_{1} = 0 \; , \;  V_{1} \propto \ast \sigma  $ (timelike).
	
	$\lambda_{2} = \sqrt{B^2 - A^2}  \; , \; V_{2} \propto \sigma + 
	\sqrt{B^2 - A^2} \theta^3 $ (spacelike).
	
	$\lambda_{3} = - \sqrt{B^2 - A^2}   \; , \; V_{3} \propto \sigma - 
	\sqrt{B^2 - A^2} \theta^3 $ (spacelike).

	\item  $B < A$ ($\sigma$ timelike)
	
	In this case we have  
	$\lambda_{1} = 0 \; , \;  V_{1} \propto \ast \sigma  $ (spacelike).
	
	$\lambda_{2} = i \sqrt{A^2 - B^2}  \; , \; V_{2} \propto \sigma + 
	i \sqrt{A^2 - B^2} \theta^3 $.
	
	$\lambda_{3} = -i \sqrt{A^2 - B^2}   \; , \; V_{3} \propto \sigma - i
	\sqrt{A^2 - B^2} \theta^3 $.

\end{enumerate} 

In order to complete the family of 1-forms that will be  used to write an
exterior system equivalent to the Einstein equations, we
introduce two new 1-forms $\beta$ and $\nu$,
 \begin{eqnarray}
 \beta & = & - (\gamma_{022} + \gamma_{033}) \theta^0 - 
 (\gamma_{122} + \gamma_{133}) \theta^1
 	\label{K4}  \\
\nu	 & = & \gamma_{010} \theta^0 + \gamma_{011} \theta^1.
 	\label{K5}
 \end{eqnarray} 
 
The interpretation of these 1-forms will be found immediately below.

 \subsection{Vanishing torsion equations}
 With our choice of tetrad and variables, the vanishing torsion 
 equations can be written as,
 
 \begin{eqnarray}
 	d \theta^0 & = &  \nu \wedge \theta^1
 	\label{vt1}  \\
 	d \theta^1 & = & \nu \wedge \theta^0
 	\label{vt2}  \\
 	d \theta^2 & = &  \alpha \wedge \theta^2 + \omega \wedge \theta^3
 	\label{vt3}  \\
 	d \theta^3 & = & (\beta - \alpha) \wedge \theta^3 + \sigma \wedge 
 	\theta^2
 	\label{vt4}
 \end{eqnarray}

The  meaning of $\nu$ and $\beta$ is now clear, since  $\nu$ is the 
 connection form in the $\theta^0-\theta^1$ subspace, while 
 $d(\theta^2 \wedge \theta^3) = \beta \wedge \theta^2 \wedge \theta^3$
 shows that $\beta$ describes the expansion of the volume in the 
 $\theta^2-\theta^3$ subspace \footnote{Note that $\beta$ is proportional to
the differential of the transitivity surface area element, as defined in
\cite{senomex} and also is proportional to the differential of the function $W$
in eq. (15.3)  in \cite{Kramer}.}.
 
 \subsection{First Bianchi identities}
 
 The first Bianchi identities are the integrability conditions for the 
 equations (\ref{vt1}-\ref{vt4}). By exterior differentiation of those 
 equations we find,
 \begin{eqnarray}
 	d \beta = 0, &  & 
 	\label{1b1}  \\
 	d \Omega + \kappa \wedge \delta  = 0, &  & 
 	\label{1b2}  \\
 	d \delta - \kappa \wedge \Omega = 0, &  & 
 	\label{1b3}  \\
 	d \kappa + \Omega \wedge \delta = 0, &  & 
 	\label{1b4}
 \end{eqnarray}
 where $\Omega \equiv \beta - 2 \alpha$, $\delta \equiv \omega + \sigma $ 
 and $ \kappa \equiv \omega - \sigma$.
 
 \subsection{Einstein field equations}
 We consider the energy-momentum tensor of a perfect fluid, $T_{\alpha 
 \nu} = (\mu + p) u_{\alpha} u_{\nu} + p g_{\alpha \nu}$ (where 
 $u_{\alpha}$ is the velocity of the fluid, $\mu$ 
 is the energy density and $p$ the pressure of the fluid), as the 
 source of the gravitational field. In appropriate units, the Einstein 
 equations read as, 
 $$R_{\alpha \nu} - \frac{1}{2} R g_{\alpha \nu} = 
 (\mu + p) u_{\alpha} u_{\nu} + p g_{\alpha \nu}$$
 or, equivalently,
 $$R_{\alpha \nu} = (\mu + p) u_{\alpha} u_{\nu} + 
 \frac{1}{2} (\mu - p) g_{\alpha \nu}.$$

After writing the Ricci tensor in the orthonormal coframe in terms of the Ricci
rotation coefficients and identifying the kinematical $1$-forms and their
derivatives, the Einstein field equations can be combined to produce the
following equivalent exterior system,\footnote{A similar calculation can be
found in \cite{tesis} for the stationary axisymmetric case.}
 \begin{eqnarray}
 	d \ast \Omega + \beta \wedge \ast \Omega + \kappa \wedge \ast 
 	\delta & =  & 0,
 	\label{fe1}  \\
 	d \ast \delta + \beta \wedge \ast \delta - \kappa \wedge \ast 
 	\Omega  & = & 0, 
 	\label{fe2}  \\
 	d \ast \beta + \beta \wedge \ast \beta  & = &  (\mu - p) u
 	\wedge \ast u,
 	\label{fe3}  \\
 	d \nu - \frac{1}{4} \beta \wedge \ast \beta + \frac{1}{4} \Omega 
 	\wedge \ast \Omega + \frac{1}{4} \delta \wedge \ast 
 	\delta  & = & - \frac{1}{2} (\mu + p) u \wedge \ast u,
 	\label{fe4}  \\
 	d \ast \tilde{\beta} + \frac{1}{2} \beta \wedge \ast 
 	\tilde{\beta} + \frac{1}{2} \Omega \wedge \ast \tilde{\Omega} + 
 	\frac{1}{2} \delta \wedge \ast \tilde{\delta} + 2 \nu \wedge \tilde{\beta}
 	 & = & - (\mu + p) u \wedge \ast \tilde{u},
 	\label{fe5}  \\
 	d  \tilde{\beta} + \frac{1}{2} \beta \wedge  
 	\tilde{\beta} + \frac{1}{2} \Omega \wedge  \tilde{\Omega} + 
 	\frac{1}{2} \delta \wedge  \tilde{\delta} + 2 \nu \wedge \ast 
 	 \tilde{\beta} & = & - (\mu + p) u \wedge  \tilde{u},
 	\label{fe6}
 \end{eqnarray}
 where the tilde operation on 1-forms in the $\theta^0-\theta^1$ 
 subspace is a 
 reflection with respect to the direction determined by $\theta^1$. If $\lambda
= 
 \lambda_{0} \theta^0 + \lambda_{1} \theta^1$, then $\tilde{\lambda} 
 \equiv \lambda_{0} \theta^0 - \lambda_{1} \theta^1$. 
 The $\ast$ operator is the Hodge dual in the 2-subspace $\theta^0-\theta^1$.
 \subsection{Integrability conditions}
 As $u$ and the other physically relevant 1-forms have vanishing Lie 
 derivative with respect to the two Killing fields, the coefficients 
 of the forms we have considered are independent of the variables in 
 the $\theta^2-\theta^3$ subspace. The integrability conditions for 
 the system of equations that  we have considered above are either trivial 
 or already incorporated in the system, except for the contracted 
 Bianchi identity,
 $$T^{\mu \nu} \; _{; \nu} = 0,$$
 which in our case reduces to,
 \begin{eqnarray}
 	d u + \frac{1}{\mu + p} d p \wedge u & = & 0,
 	\label{bi1}  \\
 	d \ast u + (\beta + \frac{1}{\mu + p} d \mu ) \wedge \ast u & = & 
 	0,
 	\label{bi2}
 \end{eqnarray}
and therefore the fluid is irrotational.

 We also have to take into account the following constraint indicating 
 that the velocity of the fluid is a unit timelike vector:
 $$ u \wedge \ast u = - \theta^0 \wedge \theta^1.$$
 
 \subsection{Simplification of the equations}
 Recall that there is a remaining freedom of choice consisting in 
 a rotation of $\theta^0$ and $\theta^1$ in their own plane, as well 
 as a rotation of $\theta^2$ and $\theta^2$ in their own plane. This
 freedom can be used to simplify the differential system introduced 
 above.
 \subsubsection{Gauge freedom in the $\theta^2-\theta^3$ subspace}
 Under a rotation through an angle $\varphi$,
 \begin{equation}
  \left(	\begin{array}{c}
  		\hat{\theta^2}  \\
  		\hat{\theta^3}
  	\end{array} \right) = 
  	\left(\begin{array}{cc}
  		\cos(\varphi) &  \sin(\varphi)  \\
  		-\sin(\varphi) & \cos(\varphi)
  	\end{array} \right) 
  	\left(\begin{array}{c}
  	   \theta^2  \\
  		\theta^3
  	\end{array}\right)
  	\label{rot23}
  \end{equation} 
  the kinematical 1-forms transform as,
  \begin{eqnarray}
  	\hat{\beta} & = & \beta,
  	\label{t21}  \\
  	\hat{\kappa} & = &  \kappa + 2 d \varphi, 
  	\label{t22}  
  \end{eqnarray}
 \begin{equation}
  \left(	\begin{array}{c}
  		\hat{\Omega}  \\
  		\hat{\delta}
  	\end{array} \right) = 
  	\left(\begin{array}{cc}
  		\cos(2 \varphi) &  - \sin(2 \varphi)  \\
  		\sin(2 \varphi) & \cos(2 \varphi)
  	\end{array} \right) 
  	\left(\begin{array}{c}
  	   \Omega  \\
  		\delta
  	\end{array}\right).
  	\label{t23}
  \end{equation} 
Without loss of generality, we can use this freedom in order to 
impose, for instance,
$$\hat{\kappa} = \epsilon \hat{\delta} \; \; ; \; \;  (\epsilon^2 =1), $$
(note that $\kappa = \delta \Leftrightarrow  \sigma = 0$ and 
$\kappa = - \delta \Leftrightarrow  \omega = 0$) or,
$$\hat{\kappa} = \epsilon \hat{\Omega} \; \; ; \; \;  (\epsilon^2 =1) $$
as the integrability condition for those equations derived from 
eq.(\ref{t22}) are identically satisfied using the first Bianchi 
identities. A more  general gauge fixing condition is,
$$\hat{\kappa} = \cos(2 \varphi_{0}) \hat{\Omega} - 
\sin(2 \varphi_{0}) \hat{\delta} \; \; ; \; \;  \varphi_{0} = {\rm 
const.}, $$ 
and in the same way as in the previous cases the integrability 
conditions are satisfied using eqs. (\ref{1b1}-\ref{1b4}).

It is important, in order to characterize the solutions of the 
equations, to know if a given solution is ``diagonal'' or not. In an 
adapted gauge this is equivalent to $\delta = 0 = \kappa$. In a non-adapted 
gauge we get $d \kappa = 0$ and $\Omega \wedge \delta = 0$. 
In other words if we have a solution satisfying the previous 
equations, with an adequate transformation, the metric can be written in 
a diagonal form.  
\subsubsection{Gauge freedom in the $\theta^0-\theta^1$ subspace} 
The freedom in the $\theta^0-\theta^1$ subspace can be used to align 
 $\theta^0$ with a timelike kinematical form or $\theta^1$ with a 
 spacelike kinematical form; a good candidate seems to be $u$ but  
 a careful analysis shows that $\beta$ is more useful. In this way we have 
 to separate three different cases:

 \noindent 1) {\bf $\beta$ null}
 	
 	If $\beta$ is a null form then we have,
 	$$\ast \beta = \epsilon \beta  \; \; ; \; \;  \epsilon^2 = 1. $$
 	
Using this relation in eq. (\ref{fe5}) and adding eq. (\ref{fe5}) with
 	eq. (\ref{fe6}) we get 
 	$$\Omega \wedge ( \tilde{\Omega} +  \epsilon \ast \tilde{\Omega}) + 
 	  \delta \wedge ( \tilde{\delta} +  \epsilon \ast \tilde{\delta}) +
 	  4 p u \wedge ( \tilde{u} +  \epsilon \ast \tilde{u}) = 0,
 	$$
    and introducing the relations,
    \begin{eqnarray*}
    	\Omega & = & \Omega_{0} \theta^0 + \Omega_{1} \theta^1, \\
    	\delta & = & \delta_{0} \theta^0 + \delta_{1} \theta^1, \\
    	u & = & u_{0} \theta^0 + u_{1} \theta^1, 
    \end{eqnarray*}
    in the previous equation we get,
    $$ (\Omega_{0} + \epsilon \Omega_{1})^2 + 
     (\delta_{0} + \epsilon \delta_{1})^2 +
    4 p (u_{0} + \epsilon u_{1})^2 = 0. $$

    As a consequence if $p > 0$ we get that $\Omega$, $\delta$, and 
    $u$ are null forms. Hence,  
    
    \centerline{{\sl  if $\beta$ null then either $p \leq 0$ or $u$ is null. }}
   
    The cases with $p<0$ or $u$ null are not very physical for a perfect fluid
    and we will 
    not consider them here. The only remaining case is  $p=0$ but from 
    (\ref{fe3}) and the fact that $\beta$ is null we get $\mu = 0$ 
    and then the solution represents a vacuum spacetime.

\noindent 2)  {\bf $\beta$ timelike}
 	
 	If $\beta$ is timelike  we can align $\theta^0$ with it. With this 
 	choice, the equations in [\ref{vt1}-\ref{bi2}] to be solved reduce to the 
 	system composed by: 
 	
 	$$\beta = e^{-Q} \theta^0$$
  	$$u\wedge \ast u = e^{2 Q} \beta \wedge \ast \beta, $$
plus the last two vanishing torsion equations, (\ref{vt3}) and (\ref{vt4}), the 
first Bianchi identities, [\ref{1b1}-\ref{1b4}], the first three Einstein
equations, [\ref{fe1}-\ref{fe3}], the integrability conditions, (\ref{bi1}) and
(\ref{bi2}), and, finally,
  	    \begin{eqnarray}
    	d Q & = &\frac{3}{4} \beta - \frac{1}{2} e^{2 Q} \{ (\mu - p) 
    	\beta + (\mu + p) [ <\beta, u> \ u + < \beta, \ast u> \ \ast 
    	u] \label{dqt} \\
    & &	  + \frac{1}{2} [ <\beta, \Omega> \ \Omega + < \beta, \ast \Omega> \ 
    	\ast \Omega]
    	+ \frac{1}{2} [ <\beta, \delta> \ \delta + < \beta, \ast \delta> \ 
    	\ast \delta]
    	\}, \nonumber
     \end{eqnarray}
   where $<,>$ stands for the scalar product and $< \theta^0,\theta^0> = -1$, $<
\theta^1,\theta^1> = 1$ , and 
   $< \theta^0,\theta^1> = 0 = < \theta^1,\theta^0>$.
   It is important to notice that the connection  $\nu$ can be solved 
   algebraically from (\ref{fe5}) and  (\ref{fe6}) and the $\nu$ thus 
   obtained 
   satisfies (\ref{fe4}) identically, using the remaining equations of the
   system. Thus, we can forget about $\nu$, which 
   may be obtained trivially after we have solved the system presented above. 
   A useful expression for $\nu$ is the following,
   $$\nu = \ast d Q - [1 - (\mu - p) e^{2 Q}] \ast \beta .$$

   Also, it has to be noticed that for a generic 1-form $\alpha$ in the
$\theta^0-\theta^1$-subspace we can 
   write,
   $$\alpha = - <\alpha, u> \ u + < \alpha, \ast u> \ 
    	\ast u, $$
   and as a consequence,
   $$\alpha_{Ru} =  <\alpha, u> \ u + < \alpha, \ast u> \ 
    	\ast u, $$ 
   can be interpreted as a reflection of $\alpha$ with respect to $u$. 
   An analogous interpretation is possible for the similar 
   expression changing $u$ by $\Omega$ (or $\delta$	). Finally let us 
   note that the integrability condition for the equation (\ref{dqt}) is 
   identically satisfied using the rest of the system.
   
\noindent 3) {\bf   $\beta$ spacelike}

If $\beta$ is spacelike we can align $\theta^1$ with it. In a similar way as in
the timelike case, the system can be reduced to [\ref{vt3}-\ref{fe3},
\ref{bi1}-\ref{bi2}] and,

 \begin{eqnarray*} 
 \beta & = & e^{-Q} \theta^1 \\
  u	\wedge \ast u & = & - e^{2 Q} \beta \wedge \ast \beta 
  	\end{eqnarray*}
    \begin{eqnarray}
    	d Q & = &\frac{3}{4} \beta + \frac{1}{2} e^{2 Q} \{ (\mu - p) 
    	\beta + (\mu + p) [ <\beta, u> \ u + < \beta, \ast u> \ \ast 
    	u]  \label{dqs} \\
    & &	  + \frac{1}{2} [ <\beta, \Omega> \ \Omega + < \beta, \ast \Omega> \ 
    	\ast \Omega]
    	+ \frac{1}{2} [ <\beta, \delta> \ \delta + < \beta, \ast \delta> \ 
    	\ast \delta]
    	\}. \nonumber 
     \end{eqnarray}

   The reasoning for timelike $\beta$ is valid for the spacelike case. The
$1$-form $\nu$ can be written as,
   $$\nu = \ast d Q - [1 + (\mu - p) e^{2 Q}] \ast \beta .$$
   Again the integrability condition for the equation 
   (\ref{dqs}) is identically satisfied.

\noindent {\bf Components of the differential of $Q$ in the light-cone and
positivity}

The spacelike or timelike character of $\beta$ is essential in 
many senses \cite{Kramer, senomex}. Here we present another one related with 
the gradient of the modulus of $\beta$, parametrized by $e^{2 Q}$. 

Hence, let us study the components of the differential of $Q$ in the light-cone,
that is, along the null geodesic directions on the $\theta^0-\theta^1$-subspace:

\noindent {\bf   $\beta$ timelike}

We have,
\begin{eqnarray*}
        \beta  & = & e^{-Q} \theta^0 \\
    	\Omega & = & \Omega_{0} \theta^0 + \Omega_{1} \theta^1 \\
    	\delta & = & \delta_{0} \theta^0 + \delta_{1} \theta^1 \\
    	u & = & u_{0} \theta^0 + u_{1} \theta^1 \\
    	d Q & = & \partial_{0} Q \theta^0 + \partial_{1} Q \theta^1 \\
    \end{eqnarray*}
and then we get,
\begin{eqnarray*}
	 e^{-Q} (\partial_{0} \pm \partial_{1}) Q& = & \frac{3}{4}  e^{-2 Q} - 
	 \frac{1}{2} (\mu - p) + \frac{1}{2} (\mu + p) 
	   (u_{0} \pm u_{1})^2  + \\
	   & & \frac{1}{4} (\Omega_{0} \pm \Omega_{1})^2 +
	  \frac{1}{4} (\delta_{0} \pm \delta_{1})^2 
\end{eqnarray*}

\noindent {\bf   $\beta$ spacelike}

The only new equation is,
\begin{eqnarray*}
        \beta  & = & e^{-Q} \theta^1
    \end{eqnarray*}
and then the new components are,
\begin{eqnarray*}
	 e^{-Q} (\partial_{1} \pm \partial_{0}) Q& = & \frac{3}{4}  e^{-2 Q} + 
	 \frac{1}{2} (\mu - p) + \frac{1}{2} (\mu + p) 
	   (u_{0} \pm u_{1})^2  + \\
	    & & \frac{1}{4} (\Omega_{0} \pm \Omega_{1})^2 +
	  \frac{1}{4} (\delta_{0} \pm \delta_{1})^2 
\end{eqnarray*}

We can see that the right-hand side of the
equation for
$(\partial_1 \pm \partial_0) Q$ is positive in the spacelike case
 (when the energy condition $\mu 
\geq |p|$ is satisfied)  and has not definite sign in the timelike case.
Therefore in the spacelike case the transitivity cylinders are not trapped
surfaces.

\section{An example}

In order to show how the formalism works let us  derive the
nonsingular solution in \cite{leo}.
This nonsingular cosmology  is diagonal, therefore 
we can choose a gauge such that 
$\omega=\sigma=0$. The $1$-form
$\beta$  is spacelike as in every other regular inhomogeneous spacetime 
and $\theta^1$ is chosen parallel to it as indicated in section 3.5.2.

Concerning the fluid, two Ans\"atze are made. The velocity of the fluid 
is taken to be $-\theta^0$ and the fluid is stiff, that is, $p=\mu$. 
Then we get that $u = e^{Q} \ast \beta$. Introducing this expression in 
 (\ref{bi1}) and (\ref{bi2}) the resulting equations imply,
\begin{equation}
dQ=-\frac{1}{2}d\ln\mu+\beta.\label{con}
\end{equation}

The equations are written using a $1$-form basis formed by the closed 
forms, $\alpha$ and $\beta$. Locally this means that we can write,
\begin{equation}
\alpha=du,\qquad \beta=dv,
\end{equation} 
and therefore  (\ref{con}) can be integrated to yield,
\begin{equation}
\mu e^{2\,Q}=a^2e^{2\,v},\end{equation}
where $a$ is a constant.

Now we have to solve  (\ref{fe1}) and (\ref{fe2}) (note 
that the only remaining equation (\ref{dqs}) can be integrated by  
line integrals, when we have solved the rest of the equations, as its 
integrability condition is identically satisfied using the rest of the 
equations of the system). Then,
we introduce two functions $f,g$ of the chosen variables $u,v$ to 
express the Hodge dual of
$\beta$ as,
\begin{equation}
\ast dv=\frac{du+f\,dv}{g},\end{equation}
that allows us to complete the dual  of our $1$-form local basis,
\begin{equation}
\ast du=\frac{-f\,du+(g^2-f^2)\,dv}{g},
\end{equation}
taking into account the properties of the Hodge dual. 

Now  (\ref{fe1}) and (\ref{fe2}) are equivalent to the 
following two partial differential equations,
\begin{equation}
g\frac {\partial f}{\partial u}-f{
\frac {\partial g}{\partial u}}+{\frac {\partial g}{\partial v}}-g
=0,\label{short}
\end{equation}
\begin{equation}
-{\frac {\partial f}{\partial v}}+f{\frac {\partial 
f}{\partial u}}-g{\frac {\partial g}{\partial u}}=0,\label{int}
\end{equation}

A simple solution of the system of equations can be obtained by requiring 
that $f$ be a function of $v$. Then  $g$ can be integrated from 
(\ref{int}),
\begin{equation}
g=\sqrt{h-2\,u{\frac {df}{dv}}},
\end{equation}
where $h$ is a function of $v$. Introducing this expression for $g$ in 
(\ref{short}) we get,
\begin{equation}
(-\frac{d^2 f}{d v^2} + 2 \frac{d f}{d v}) u + \frac{1}{2} \frac{d 
h}{d v} - h + f \frac{d f}{d v} = 0,
\end{equation}
which can be easily integrated to get,
\begin{equation}
f=p+q\,e^{2\,v},
\end{equation}
\begin{equation}
h=-2\,{q}^{2}{e^{4\,v}}-4\,{e^{2\,v}}pqv+wq{e^{2\,v}},
\end{equation}
where $p$, $q$, and $w$ are constants.
This allows us to write down the function $g$,
\begin{equation}
g=\sqrt {-2\,{q}^{2}{e^{4\,v}}-4\,{e^{2\,v}}pqv+wq{e^{2\,v}}-4\,qu{e^{2
\,v}}}.
\end{equation}

The last equation to be integrated is (\ref{dqs}). Using the 
expression for $Q$ that has been obtained previously, it can be written as
follows,
\begin{equation}
\frac{d \mu}{\mu} = (2 f^2 - 2 g^2 - 2 a^2 e^{2 v}) dv + (2 + 4 f) du,
\end{equation}
which allows us to obtain the following expression for the fluid energy 
density,
\begin{equation}
\mu=k\,\exp\left(2\,{p}^{2}v+\frac{3}{2}\,{q}^{2}{e^{4\,v}}+
4\,{e^{2\,v}}pqv-wq{e^{2\,v}}-a^{2}{e^{2\,v}}
+2\,u+4\,pu+4\,qu{e^{2\,v}}\right),
\end{equation}
where the constant $k$ will be taken to be $a^2$ without loss of generality.

The metric can be written in a simple form if we perform a change of 
variables and the coordinate $v$ is taken to be $\ln r$. If $t$ is 
chosen accordingly so that the coordinates are isotropic,
\begin{equation}
v=\ln r,\qquad u=-\frac{1}{2}\,q{r}^{2}-p\ln r-q{t}^{2},
\end{equation}
the fluid density takes the form,

\begin{equation}
\mu=a^2\,{r}^{-2p(p+1)}\exp\left(-\frac{1}{2}\,{q}^{2}{r}^{4}-(
a^2+q+2\,pq){r}^{2}-2\,q{t}^{2}-4\,pq{
t}^{2}-4\,q^2{t}^{2}{r}^{2}\right),
\end{equation}
after eliminating unnecessary parameters.

The metric functions  or the fluid density are singular unless $p=0,-1$. 
Both cases are indeed the
same solution just changing $q$ for $-q$. In \cite{leo}, $q$ is named 
$\beta$ and $a^2$ is $\alpha$.

Note that, in principle, this metric could also have been  obtained by the
generation algorithm due to Wainwright, Ince and Marsh \cite{Wain}.

\section{Geodesic Completeness}

In this section we shall attempt to give a sufficient condition for 
a diagonal cylindrical $G_2$-cosmology to
be non-singular. Once one encounters a cosmological model with 
regular curvature invariants it usually
takes lengthy and tedious calculations to prove that the metric 
is causally geodesically complete. The
condition we state here shall be large enough to comprise most 
known cylindrical non-singular cosmological
models.

For diagonal cylindrical $G_2$-cosmology we can choose a gauge where 
$\omega = 0 = \sigma$ (note that there is a residual gauge, a 
rotation with $\varphi =$ const.). The equations (\ref{1b1}) and (\ref{1b2}) 
can be integrated obtaining,
\begin{equation}
	\alpha = - d f  \;\; , \;\; \beta = \frac{d \rho}{\rho}.
	\label{intab}
\end{equation}

Using these expressions for $\alpha$ and $\beta$, as well as the fact 
that $\omega = 0 = \sigma$, we can integrate equations (\ref{vt3}) 
and (\ref{vt4}) to get,
\begin{eqnarray*}
	\theta^2 & = & e^{-f} d z, \\
	\theta^3 & = & \rho e^{f} d \phi.
	\label{intt23}
\end{eqnarray*}

All known non-singular diagonal cylindrical $G_2$-cosmologies 
have a spacelike $\beta$. Hence, following the method described in 
section 3.5.2, we align $\theta^1$ with $\beta$,
\begin{equation}
  \theta^1 = e^{Q} \frac{d \rho}{\rho}.
	\label{tet1}
\end{equation}

As the cosmological fluid is irrotational, 
the velocity of the fluid can be written as,
$$u = e^g d t,$$
using isotropic coordinates $t$ and $r$ such that,
  \begin{equation}
  	\ast dt = - dr \;\; {\rm and} \;\; \ast dr = - dt,
  	\label{dual}
  \end{equation}
that provides the following expression for $\theta^0=- \ast \theta^1$,
  \begin{equation}
  	\theta^0 = e^{Q} (\frac{\rho_{r}}{\rho} dt + \frac{\rho_{t}}{\rho} 
  	dr).
  	\label{tet0}
  \end{equation}  
The constraint $u \wedge \ast u = - e^{2 Q} \beta \wedge \ast \beta$ 
imposes a relation between $g$ and $Q$,

\begin{equation}
  	e^{2 Q} = \frac{\rho^2 e^{2 g}}{\rho_{r}^2 - \rho_{t}^2},
  	\label{qgcons}
\end{equation} 

Using the previous results  the metric takes the following form,
\begin{eqnarray}
ds^2=e^{2\,g(t,r)}\left\{-dt^2+dr^2\right\}+\rho^2(t,r)e^{2\,f(t,r)}d\phi^2
+e^{-2\,f(t,r)}dz^2.\label{metric}
\end{eqnarray}

Hence, the calculations will be performed in a chart where $t$ and $r$ are 
isotropic coordinates for the subspace
orthogonal to the Killing orbits and $\phi$ and $z$ are coordinates 
adapted to the Killing fields. The
usual ranges for these coordinates are chosen,

\begin{eqnarray}
-\infty<t,z<\infty,\quad 0<r<\infty,\quad 0<\phi<2\,\pi.
\end{eqnarray}

We shall assume from the beginning that the metric functions $f,g,\rho$ are $C^2$ and that $\rho$ is 
positive. Certainly a $C^2$ requirement is needed in order to avoid a singular Riemann tensor, but for
geodesic completeness $C^1$ would be enough since just the affine connection is involved in the equations.

Following \cite{Kramer}, if we are to have a regular symmetry axis on
the locus where $\Delta=g(\xi,\xi)=0$, then we have to impose that,

\begin{equation}
\lim_{\rho\to
0}\frac{g({\rm grad}\,\Delta,{\rm grad}\,\Delta)}{4\,\Delta}=
1,\label{reg} \end{equation}
on approaching the axis. In our case this means, if the metric functions $f,g$ are $C^1$ at the axis,

\begin{eqnarray}
\lim_{\rho\rightarrow
0}e^{2\,\left\{f(t,r)-g(t,r)\right\}}\left\{\rho_r(t,r)^2-
\rho_t(t,r)^2\right\}=1.
\end{eqnarray}

From now on we shall denote the derivatives with respect to $r$ and $t$ by
subscripts. 

There is no loss of generality in imposing that the axis is located on $r=0$, since we are always free to
choose a different set of isotropic coordinates $T,R$ by performing a transformation,
\begin{eqnarray} T_{t}=R_r,\qquad T_{r}=R_{t},
\end{eqnarray}
which amounts to take a solution of the one-dimensional wave equation,
\begin{eqnarray} R_{tt}-R_{rr}=0.
\end{eqnarray}
with a boundary condition $R=0$ on the axis. This problem is underdetermined, since no initial condition
has been imposed. For instance, one could take $R=r$ and $R_t=0$  at $t=0$.

 In order to determine whether the metric is geodesically complete, we shall write the
expressions for the geodesic equations,
\begin{equation} 
\ddot x^i+\Gamma^i_{jk}\dot x^j\dot x^k = 0,\end{equation}
where the dot denotes derivative with respect to the affine parameter of the geodesic.

 In principle we would
have to write a second order system of four differential equations, but two of them become first order due
to the existence of isometries. If $u=(\dot t,\dot r,\dot\phi,\dot z)$ is the $4$-velocity of the geodesic
and
$\xi$ is a Killing field then $p=\xi\cdot u$ is a constant of geodesic motion. The following quantities are
then conserved along geodesics,

\begin{eqnarray}
L=\rho^{2}(t,r)e^{2\,f(t,r)}\dot\phi,
\end{eqnarray}

\begin{eqnarray}
P=e^{-2\,f(t,r)}\dot z,
\end{eqnarray}
respectively the angular momentum  and the $z$ component of the linear
momentum of a unit mass test particle in free fall.

 There is also a conserved quantity,
$\delta$, which takes the value zero for lightlike, one for timelike and minus one for spacelike
geodesics. This quantity arises from the fact that the $4$-velocity $u$ is normalized when an affine
parametrization is used. We shall consider just future causal geodesics,

\begin{eqnarray}
\delta=e^{2\,g(t,r)}\left\{\dot
t^2-\dot r^2\right\}-L^2\rho^{-2}(t,r)e^{-2\,f(t,r)}-P^2e^{2\,f(t,r)}.\label{delta}
\end{eqnarray}

The remaining second order equations read,

\begin{eqnarray}
\ddot t+g_t(t,r)\dot t^{2}
+2\,g_r(t,r){
\dot t}{\dot r}+g_t(t,r){\dot r}
^{2}-{P}^{2}{e^{2\,\left\{f(t,r)-g(t,r)\right\}}}f_t(t,r)
+\nonumber\\+{L}^{2}\frac{e^{-2\,\left\{f(t,r)+g(t,r)\right\}}}{\rho(t,r)^{3}}\left\{\rho_t(t,r)
+\rho(t,r)f_t(t,r)\right\},
\label{eq1}
\end{eqnarray}

\begin{eqnarray}
\ddot r+g_r(t,r)\dot t^{2}
+2\,g_t(t,r){
\dot t}{\dot r}+g_r(t,r){\dot r}
^{2}+{P}^{2}{e^{2\,\left\{f(t,r)-g(t,r)\right\}}}f_r(t,r)
-\nonumber\\-{L}^{2}\frac{e^{-2\,\left\{f(t,r)+g(t,r)\right\}}}{\rho(t,r)^{3}}\left\{\rho_r(t,r)
+\rho(t,r)f_r(t,r)\right\},\label{eq2}
\end{eqnarray}
after substituting the derivatives of the cyclic coordinates $z,\phi$ for their expressions in terms of the
constants of geodesic motion. 

These equations can be written in a more compact form making use of the equation for $\delta$,
\begin{eqnarray}
\{e^{2\,g(t,r)}\dot t\}^{\cdot} -\frac{e^{-2g(t,r)}}{2}\left\{e^{2g(t,r)}\left[\delta
+P^2e^{2f(t,r)}+L^2\frac{e^{-2f(t,r)}}{
\rho^{2}(t,r)}\right]\right\}_t=0,
\label{eeq1}
\end{eqnarray}

\begin{eqnarray}
\{e^{2\,g(t,r)}\dot
r\}^{\cdot}+\frac{e^{-2g(t,r)}}{2}\left\{e^{2g(t,r)}\left[\delta
+P^2e^{2f(t,r)}+L^2\frac{e^{-2f(t,r)}}{
\rho^{2}(t,r)}\right]\right\}_r=0,\label{eeq2}
\end{eqnarray}

An important family of geodesics are lightlike radial geodesics ($L=P=\delta=0$), for which there is a
constant $k$ such that,
\begin{eqnarray}
\dot t=|\dot r|=k\,e^{-2\,g(t,r)}.
\end{eqnarray}

The derivative of the time coordinate must not grow too fast if these geodesics are complete. A sufficient
condition is achieved by imposing that $t$ does not grow faster than exponentially for large values
of $t$. This amounts to the following condition on $g$,

\begin{eqnarray}
g(t,r)\ge-\frac{1}{2}\ln|t+a|+b,\label{radial}
\end{eqnarray}
where $a$ and $b$ are constants. If this condition is fulfilled, $t$ is defined for arbitrarily large
values of the affine parameter.

The system of second order equations (\ref{eeq1}), (\ref{eeq2}) is shown to be equivalent to a system
of first order equations by introducing a new function $\xi$. Since,

\begin{equation}
\dot t^2-\dot
r^2={e^{-2g(t,r)}}\left\{\delta+L^2\frac{e^{-2\,f(t,r)}}{\rho^{2}(t,r)}+P^2e^{2\,f(t,r)}\right\}, 
\end{equation}
it is tempting to parametrize $\dot t$, $\dot r$ by means of hyperbolic functions of $\xi$,

\begin{eqnarray}
\dot t=e^{-g(t,r)}\sqrt{\delta+L^2\frac{e^{-2f(t,r)}}{\rho^{2}(t,r)}+P^2e^{2f(t,r)}}\cosh\xi(t,r),
\end{eqnarray}

\begin{eqnarray}
\dot r=e^{-g(t,r)}\sqrt{\delta+L^2\frac{e^{-2f(t,r)}}{\rho^{2}(t,r)}+P^2e^{2f(t,r)}}\sinh\xi(t,r).
\end{eqnarray}

After introducing these expressions in (\ref{eeq1}), (\ref{eeq2}), a first order equation for $\xi$
is easily obtained,

\begin{eqnarray}
\dot\xi(t,r)=-e^{-2g(t,r)}\left\{F_t(t,r)\sinh\xi(t,r)+F_r(t,r)\cosh\xi(t,r)\right\},
\end{eqnarray}

\begin{equation}
F(t,r)=e^{g(t,r)}\sqrt{\delta+L^2\frac{e^{-2f(t,r)}}{\rho^{2}(t,r)}+P^2e^{2f(t,r)}},
\end{equation}
which expands into the following expression for $\dot\xi$,

\begin{eqnarray}
&\dot\xi=-\frac{e^{-g}}{\sqrt{\delta+L^2\rho^{-2}e^{-2f}+P^2e^{2f}}}\times\nonumber\\&
\times\left\{\cosh\xi\left[\delta
g_r+L^2\frac{e^{-2f}}{\rho^2}\left(g_r-f_r-\frac{\rho_r}{\rho}\right)+P^2e^{2f}\left(g_r+f_r\right)\right]+
\right.\nonumber\\&+\left.\sinh\xi\left[\delta
g_t+L^2\frac{e^{-2f}}{\rho^2}\left(g_t-f_t-\frac{\rho_t}{\rho}\right)+P^2e^{2f}\left(g_t+f_t\right)\right]
\right\},
\end{eqnarray}
where the dependence of the functions on $t, r$ has been omitted in order to avoid a large equation.

The equivalent equation for past-directed geodesics can be also obtained,

\begin{eqnarray}
\dot\xi(t,r)=e^{-2g(t,r)}\left\{F_t(t,r)\sinh\xi(t,r)-F_r(t,r)\cosh\xi(t,r)\right\}.
\end{eqnarray}

Now we can check the equations to extract conditions for causal geodesics to be complete. No attention
needs to be paid to the equations for $\dot z$ and $\dot \phi$, since they are just quadratures to be
solved once  $t,r$ are known as functions of the affine parameter and cannot be singular unless the functions
involved in the integration are singular and we have already imposed that the metric functions
must be smooth. 

 We have to prevent the coordinates
$t,r$ from tending to infinity at a finite value of the affine parameter. Also the coordinate $r$ must not
tend to zero at a finite value of the affine parameter when the constant of geodesic motion $L$ is nonzero.
When $L$ is zero, the denominators that depend on $\rho$ disappear and the expressions for $\dot t $, $\dot
r$, $\dot
\xi$ are not singular at $r=0$.

First of all we shall exclude the possibility of having arbitrarily  large values of $|\xi|$. There are two
cases to consider,

\begin{itemize}
\item $\dot\xi>0,\ \xi>0:$ We shall impose a condition in order to prevent this situation from lasting too
long by requiring that $\dot\xi$ changes sign for large values of the time coordinate $t$. Since we may have
geodesics with any of the constants of motion equal to zero, the terms that multiply them in the
numerator of $\dot\xi$ must be treated independently. For instance, for $\delta$, we need,

\begin{equation}
0<\cosh\xi\,g_r+\sinh\xi\,g_t=\sinh\xi\,(g_r+g_t)+e^{-\xi}\,g_r,
\end{equation} 
and  similar conditions for $L$ and $P$. These conditions can be fulfilled by requiring,

\begin{eqnarray}\label{Mxi1}
\left\{
\begin{array}{l}g_r+g_t>0\\
(g-f-\ln\rho)_r+(g-f-\ln\rho)_t>0\\
(g+f)_r+(g+f)_t>0\end{array}\right.,\end{eqnarray}
for large values of $t$ and increasing $r$. And for the remaining terms that
multiply $e^{-\xi}$, we shall have further to impose that, 

\begin{eqnarray}\label{Mxi2}
\left\{
\begin{array}{l}{g_r}\\(g-f-\ln\rho)_r\\(g+f)_r\end{array}\right..
\end{eqnarray}
be either positive or at most of the same order as their respective terms in (\ref{Mxi1}) so that
$e^{-\xi}$ turns them negligible, also for large values of $t$ and increasing $r$. 

With all these requirements $\dot\xi$ becomes negative before $t$ may
develop a singularity.

\item $\dot\xi<0,\ \xi<0:$ As in the previous case, we shall prevent $\xi$ from growing unboundedly large.
Since the unfriendly denominators depending on $\rho$, that are only dangerous in this case since $r$
decreases, only appear when
$L\neq0$, we shall just require for nonzero $L$, that $\xi$ becomes positive for large 
values of $t$.
Again, by imposing such restriction on the numerator of $\dot\xi$ one finds, for large values of $t$ and
decreasing $r$, 

\begin{eqnarray}
&\delta\{g_t-g_r\}+L^2\frac{e^{-2f}}{\rho^2}\left\{(g-f-\ln\rho)_t-(g-f-\ln\rho)_r\right\}+\nonumber\\&
P^2e^{2f}\left\{
(g+f)_t-(g+f)_r\right\}>0\label{mxi1}
\end{eqnarray}
by making use  of the identity $\cosh\xi=-\sinh\xi+e^{\xi}$. Since $L\neq0$, its term
may help compensate for the others to achieve a positive numerator.  The terms that
multiply the exponential of
$\xi$, 

\begin{eqnarray}\delta
g_r+L^2\frac{e^{-2f}}{\rho^2}\left(g_r-f_r-\frac{\rho_r}{\rho}\right)+P^2e^{2f}\left(g_r+f_r\right),\label{mxi2}
\end{eqnarray}
must be either negative or at most of the same order as the term in (\ref{mxi1}) for large values
of $t$ and decreasing $r$.
\end{itemize}

Finally we shall prescribe a behaviour for $\dot t$ similar to the one used for lightlike radial geodesics
to avoid faster than exponential growth for large values of $t$,

\begin{eqnarray}
\left.\begin{array}{c}g(t,r)\\g(t,r)+f(t,r)+\ln\rho\\g(t,r)-f(t,r)\end{array}\right\}\ge-\ln|t+a|+b,\label{tt}
\end{eqnarray}
the first of which is obviously weaker than the condition already required for lightlike radial geodesics.
The constants $a,b$ need not be the same for the three inequalities.

No similar restrictions need be imposed on $\dot r$ since this derivative cannot become singular before
$\dot t$ does according to equation (\ref{delta}).

Similar results can be derived for past-directed causal geodesics just replacing the
sign of the derivatives with respect to $t$ for the opposite   in (\ref{Mxi1}), 
(\ref{mxi1}),  for small values of the time coordinate. The other conditions
remain unchanged for small values of $t$ since they where obtained for $|\dot t|$.

According to \cite{Geroch}, the spacetime is globally hyperbolic since from the derivation of the
results it follows that every lightlike geodesic intersects once and only once every hypersurface
of constant time and therefore they are Cauchy hypersurfaces. 

These results can be summarized in the following  theorem,

\begin{description}
\item[Theorem:] A cylindrically symmetric diagonal metric in the form (\ref{metric}) with $C^2$
metric functions  $f,g,\rho$ is future causally geodesically complete if conditions (\ref{radial}),
(\ref{Mxi1}), (\ref{Mxi2}), (\ref{mxi1}), (\ref{mxi2}), (\ref{tt}) are fulfilled for large values of
$t$.

In addition, the spacetime is globally hyperbolic.
\end{description}

A fact that is worthwhile mentioning is the relation for the orbit of a future-directed geodesic,

\begin{equation}
\frac{dr}{dt}=\tanh\xi.\end{equation} 

This means that, for events $(t_0,r_0)$, $(t,r)$, $r_0<r$, $t_0<t$,  on the geodesic we have,  

\begin{equation}
r\le r_0+(t-t_0),
\end{equation}
and if $r_0>r$, $t_0<t$, then,

\begin{equation}
r\ge r_0-(t-t_0).\label{bound}
\end{equation}

Furthermore,  the orbit lies above its tangent lines while 
$\xi,\dot\xi>0$ and therefore, we get a lower bound. For $r_0<r$, $t_0<t$,

\begin{equation}r\ge r_0+
\tanh\xi(t_0,r_0)\cdot(t-t_0),\end{equation}

And similarly, for $\xi,\dot\xi<0$, $r_0>r$, $t_0<t$,

\begin{equation}r\le r_0+\tanh\xi(t_0,r_0)\cdot(t-t_0),\label{rebound}\end{equation} 

\section{Examples}

In this section we shall show explicitly how the theorem works with all
 known diagonal nonsingular models.

\begin{itemize}
\item Senovilla: \cite{Seno} This is the first known nonsingular cosmological model in the literature. It
describes a universe in a radiation dominated epoch.  Its geodesic completeness is proven in \cite{Chinea}.
The metric functions for this model are,

\begin{eqnarray}
&g(t,r)=2\,\ln\cosh(a\,t)+\ln\cosh(3\,a\,r),\nonumber\\
&f(t,r)=\ln\cosh(a\,t)+\frac{1}{3}\ln\cosh(3\,a\,r),\nonumber\\ 
&\rho(t,r)=\frac{1}{3\,a}\cosh(a\,t)\sinh(3\,a\,r)\cosh^{-2/3}(3\,a\,r),
\end{eqnarray}
where $a$ is a positive constant related to the maximum of the pressure.

Since they are even functions of $t$ we need not worry about past-directed geodesics.

All three functions in (\ref{radial}),(\ref{tt}) are positive except for constants and the term
$\ln(\sinh(3ar))$ in the third function. For increasing $r$, it does not mean a problem. For
decreasing $r$, it is bounded by (\ref{bound}),

\begin{equation}
\ln(\sinh(3ar))\ge\ln(\sinh (3a|t-r_0+t_0|)),
\end{equation}

and therefore they fulfil this condition. 

The first and third conditions in (\ref{Mxi1}) are always satisfied for positive
$t$ and the same happens with the second for large values of $r$. 

The expressions in (\ref{Mxi2}) do not depend on $t$ and are always positive, except the second
one, which requires large values of $r$.

All three terms in (\ref{mxi1}) independently are positive for small values of $r$ and positive
$t$. 

Concerning (\ref{mxi2}), the term as a whole is negative for small $r$ and positive $t$.

Hence this spacetime is geodesically complete.

\item Ruiz-Senovilla: \cite{Ruiz} This family includes the previous one as a subcase for $K=1$ and
$n=3$. The metric is not written in isotropic form in the original paper, but it can be cast in
that form by the change of variable,

\begin{eqnarray}
 &r=\frac{1}{n\,a}{\displaystyle\int_1^{\cosh(n\,a\,R)}}{\displaystyle\frac{dx}{P(x)}},\nonumber\\
&P(x)=\sqrt{x^2+(K-1)\,x^{(2n-1)/n}-K},
\end{eqnarray}
where $P(x)$ is a growing function for $x>1$. We shall need for our reasoning that,

\begin{equation}
\frac{dR(r)}{dr}=\frac{P(\cosh(n\,a\,R(r)))}{\sinh(n\,a\,R(r))},
\end{equation}
which tends to a positive constant when $r$ tends to zero, to one when $r$ tends to infinity and is
always positive for positive $r$.

\begin{eqnarray}
&g(t,r)={\displaystyle\frac{1+n}{2}\,\ln\cosh(a\,t)+\frac{n-1}{2}\,\ln\cosh(n\,a\,R(r))},\nonumber\\ &
f(t,r)={\displaystyle\frac{n-1}{2}\ln\cosh(a\,t)+\frac{n-1}{2\,n}\ln\cosh(n\,a\,R(r))},\nonumber\\ &
\rho(t,r)={\displaystyle\frac{1}{n\,a\,L}\cosh(a\,t)\cosh^{(1-n)/n}(n\,a\,R(r))P(\cosh^2(n\,a\,R(r)))},\nonumber\\
&L=K-\frac{K-1}{2\,n}.
\end{eqnarray}

The  ranges of the parameters are $n\ge3$, $K>0$ and $a$ is again an arbitrary positive constant and 
 every metric function is even in the coordinate time $t$.

Every function in in (\ref{radial}),(\ref{tt}) is either positive or involves a term that behaves
as $\ln r$ for small $r$ and therefore can be bounded like in the previous example.

The expressions in (\ref{Mxi1}) are all positive for large $r$ and positive  $t$ as it happened in
the Senovilla subcase. The first and third one are positive for all $r$.

The radial derivatives in  (\ref{Mxi2}) are again positive for large values of $r$. The first and
third one are positive for all $r$.

Again the three terms  in (\ref{mxi1}) are positive for large 
$t$ and small  values of $r$. 

Finally, the angular momentum term in (\ref{mxi2}) is unboundedly negative for small $r$ while the
others are bounded and therefore the whole term is negative for small values of the radial
coordinate.

Hence the whole set of conditions is fulfilled and the family is geodesically complete.

\item Mars: \cite{Sep} This is a family of nonsingular models with no 
equation of state and depending on two
parameters $a$, related to the maximum density of perfect fluid, and $c\ge
3/4$,\footnote{Note that there is a missprint in the original paper.}  

\begin{eqnarray}
&g(t,r)=-{\displaystyle\frac{1}{2}}\,\ln\cosh(2\,a\,t)+\frac{c}{2}\cosh^2(2\,a\,t)\sinh^2(2\,a\,r),\nonumber\\ 
&f(t,r)={\displaystyle\frac{1}{2}}\,\ln\cosh(2\,a\,t)-\ln\cosh(a\,r),\nonumber\\&
\rho(t,r)={\displaystyle\frac{1}{2\,a}\frac{\sinh(2\,a\,r)}{\cosh(2\,a\,t)}}.
\end{eqnarray}

Again the metric functions are even in the time coordinate $t$.

The negative terms in (\ref{radial}), (\ref{tt}) are counteracted by the exponential growth of the
$c$ terms for large $r$ and $t$.

Concerning (\ref{Mxi1}), the three expressions are positive for positive $t$ except for bounded terms
in $t$ and $r$.

The same sort of behaviour is to be found in (\ref{Mxi2}) but for all values of $t$.

The terms in the expression of (\ref{mxi1}) are all bounded or tend to zero except for
$2a\coth(2ra)$. Therefore, this term is positive for small $r$. The reasoning why
the terms on
$\sinh(ar)$ tend to zero faster than the ones with $\cosh(at)$ is explained in the next example.

Finally, the expression in (\ref{mxi2}) is negative for small values of $r$ since every term tends
to zero or is bounded except for the $-2a\coth(2ra)$ term. 

Therefore, this family is also geodesically complete.

\item Fern\'andez-Jambrina: This non-separable metric, that has been obtained
in section 4, corresponds to a spacetime filled with a stiff perfect fluid. Both
parameters,
$\alpha$ and
$\beta$, are positive. $\alpha$ is related to the maximum of the density of the
fluid. The metric functions are, 

\begin{eqnarray}
&g(t,r)=\frac{1}{4}\beta^2\,r^4+\frac{\alpha+\beta}{2}\,r^2+\beta\,t^2+2\,\beta^2\,r^2\,t^2,\nonumber\\
&f(t,r)=\frac{\beta}{2}(r^2+2\,t^2),\qquad \rho(t,r)=r,
\end{eqnarray}
and again they are even functions of time.

The three functions in (\ref{radial}),(\ref{tt}) are positive, except for a constant and a term $\ln
r$ to which applies the same reasoning used for Senovilla's metric  and so this
condition is trivially satisfied.

The first and third expressions in (\ref{Mxi1}) are positive for positive values of $t$. The
second expression requires in addition large values of $r$.

The same happens for  (\ref{Mxi2}), regardless this time of the sign of $t$.

The terms in (\ref{mxi1}) are all either positive and growing linearly with $t$ or as $1/r$ or
tending to zero. Note  that the terms polynomic in $r$ tend to zero although they may be multiplied
by powers of $t$ since according to (\ref{rebound}) $r$ decreases to zero
faster than
$t$ grows to infinity.

Finally, all of the terms in (\ref{mxi2}) tend to zero except $-1/r$, which makes the expression
negative for small values of $r$.

Therefore these spacetimes are geodesically complete.

\section{Summary}
In this paper a new approach for $G_2$ inhomogeneous cosmological models has
been derived. The description is grounded on  kinematical properties of
spacelike congruences that have been encoded in several 1-forms. The exterior
differential system, equivalent to Einstein equations, has been
submitted to different kinds of simplifications. Based in this formalism
several consequences are drawn. It is shown that when  the
transitivity surfaces  are null a perfect fluid is not admissible. Also, when
these surfaces are spacelike they are not trapped sets. 

As an application to non-singular cosmological models a solution is reobtained
within the formalism. A theorem is written stating a sufficient condition for
diagonal models to be singularity-free and globally hyperbolic. All known
non-singular diagonal perfect fluid spacetimes fit in this framework.

\vspace{0.3cm}
\noindent{\em The present work has been supported by Direcci\'on General de
Ense\~nanza Superior Project PB95-0371. The authors wish to thank
 F. J. Chinea and M. J. Pareja for valuable discussions.}

\end{itemize}

\end{document}